\documentclass[12pt]{article}
\usepackage{epsfig} \usepackage{amssymb} \usepackage{amsfonts}
\def\ea{{\it et al.\,}} \def\eg{{\it e.g.,\,}} \def\ie{{\it i.e.,\,}}
\def\eV{{\rm e\kern-0.12em V}} \def\GeV{{\rm G}\eV} 
\def\msbar{\relax\ifmmode\overline{\rm MS}\else{$\overline{\rm MS}${ }}\fi}
\def\albar{\relax\ifmmode{\bar{\alpha}}\else{$\,\bar{\alpha}${ }}\fi}
\def\gbar{\relax\ifmmode{\bar{g}}\else{$\,\bar{g}${ }}\fi}
\def\albarQ{\relax\ifmmode{\bar{\alpha}(Q^2)}\else{$\albar(Q^2)${ }}\fi}
\def\albarsf{\relax\ifmmode{\bar{\alpha}_{\rm SF}}\else{$\,\bar{\alpha}_{\rm SF}${ }}\fi}
\def\albarsQ{\relax\ifmmode{\bar{\alpha}_s(Q^2)}\else{$\albar_s(Q^2)${ }}\fi}
\def\alphan{\relax\ifmmode{\alpha_{\rm an}}\else{$\,\alpha_{\rm an}${ }}\fi}
\def\albars{\relax\ifmmode{\bar{\alpha}_s}\else{$\,\bar{\alpha}_s${ }}\fi}
\def\tildal{\relax\ifmmode{\tilde{\alpha}}\else{$\,\tilde{\alpha}${ }}\fi}
\def\as{\relax\ifmmode\,\alpha_s\,\else{$\,\alpha_s\,${ }}\fi}
\def\asQ{\relax\ifmmode{\/\bar{\alpha}_s(Q^2)}\else{$\/\albars(Q^2)${ }}\fi}
\newcommand{\beglab}{\begin{equation}\label}
\newcommand{\beq}{\begin{equation}} \newcommand{\eeq}{\end{equation}}
\begin{document}
\begin{flushright}
 To appear in TMP vol.132, No 3 (Sept.2002)

\end{flushright} \vspace{15mm}

\begin{center}
{\large\sf On the QCD coupling behavior in the
           infrared region} \\
\medskip

D.V. Shirkov \\
{\it Bogoliubov Laboratory, JINR, 141980 Dubna, Russia\\
e-address: shirkovd@thsun1.jinr.ru} \end{center} \smallskip

{\small  The summary of nonperturbative results for the QCD
invariant coupling \albars obtained by lattice simulations for
functional integral and by solution of approximate
Dyson--Schwinger equations reveals a puzzling variety of IR
behavior of \asQ even on a qualitative level. This,
in turn, rises a question of correspondence between the
results obtained so far by different groups.\par
  We analyze this issue in terms of mass-dependent
coupling constant transformations and conclude that
the question of the IR behavior of effective QCD
coupling and of propagators is not a well--defined
one and needs to be more specified.} \bigskip

\tableofcontents
\newpage
\section{Introduction}

This paper is devoted to the issue of QCD
invariant (effective) coupling function \albar
behavior in the infrared (IR) region. \par
 The notion of invariant coupling is the central one in current practice
of quantum--field applications of the renormalization group (RG)
method. In QFT, the very existence of RG is generically connected
with Dyson's finite renormalization transformations. The invariant
coupling (IC) --- or invariant charge --- has been introduced (see
pioneer papers \cite{bsh55rg1,bsh55rg2, bsh56nc}, as well as
\cite{rg-kniga}) via a product of finite Dyson's renormalization
factors $z_i\,,$ the product that is invariant with respect to the
mentioned transformations, without any reference to weak coupling
or ultraviolet (UV) limits. \par

 By expressing $z_i\,$ via values of renormalized QFT functions one
usually obtains an expression for IC \albarQ as a function of one
argument $\,Q=\sqrt{Q^2={\bf Q^2}-Q^2_0}\,$ --- momentum transfer
(or reference momentum) --- in terms of a product of the coupling
constant, and the vertex and propagator Lorentz--invariant
amplitudes taken in the momentum representation\footnote{See, \eg,
eqs.(\ref{a-tub}), (\ref{a-par}) and (\ref{a-desy}) below.} in
some particular renormalization scheme. \par

  Such definitions have been introduced and then used in the
perturbation case for the UV and IR asymptotics in the mid-50s. In
current practice, \albarQ is usually employed in the UV limit only.
The massless minimal subtraction renormalization scheme  \msbar
turns out to be the most convenient one and, hence, more popular in the
 UV analysis. \par

Meanwhile, the mentioned definitions are valid in a more general,
mass dependent, case --- see
refs.\cite{bsh55rg1,bsh56nc,rg-kniga,92-3}. In the gauge QFT
models containing several vertices with the same coupling constant
this mass dependence, in turn, yields a specific {\it vertex
dependence}, \ie the dependence on a particular dressed vertex
chosen for the IC defining. In QCD, for instance, one can use
three--gluon, gluon--ghost, four--gluon and various gluon--quark
vertices. In the massless \msbar scheme, all ICs constructed with
the use of different vertices are the same. However, in the
massive case, they could differ essentially \cite{95-1} in the
$\,Q\lesssim m\,$ region. \par

 Here, the comment on the difference between massless and mass--dependent
subtraction schemes and related RG solutions is in order. For example,
there exist two variants of the \msbar scheme. In the common one, for
the QFT model with one coupling constant $\,g\,,$ with the help of
massless counter--terms one constructs a beta--function $\,\beta(g)\,,$
the function of one argument. The RG equation solving for IC results in
expression $\,\gbar=f(\ell,g)\,$ depending on the logarithm
$\,\ell=\ln(Q^2/\mu^2)\,$ and describing the UV asymptotic behavior.
Meanwhile, in a more general formalism, the RG generators (beta and
gamma functions) are defined with the help of the \msbar renormalized,
mass--dependent approximate expressions. They are functions of two
dimensionless arguments $\,Q^2/m^2\,$ and $\,g\,.$ The RG equations
solving produces expressions with the additional ``massive" argument
$\,m^2/\mu^2\,.$ The general expression for IC is now of the form
$\,\gbar(Q^2/\mu^2,m^2/\mu^2,g)\,.$ The mass dependence is essential
in the IR domain. This algorithm was used in QED\cite{bsh55rg2,tolja}
for describing the IR singularity of the electron propagator.\par

 This mass dependence, in our opinion, reveals itself in the current
analysis of the IR properties in QCD, in particular, at lattice
calculation of the functional integral and in solving  the
truncated Dyson--Schwinger equations (DSE). Various groups of
researchers use different definitions of coupling constants and IC
in the IR region. For instance, in the IC defining the T\"ubingen
group uses\cite{as01} the gluon--ghost vertex, the Paris group
\cite{b1-171,b1-278} --- three--gluon one, while the
``Transoceanic team" \cite{skw01,sk02} --- the gluon--quark
vertex. \par

 Section 2 contains a short overview of different groups' results
for \albars obtained  by solution of truncated DSE as well as by
lattice simulations in the IR region. This summary reveals a rather
wide variety of IR behaviors. \par

 In Section 3, we remind the origin of the IC \asQ notion and note
that it is free of any relation with the weak coupling or UV
massless limits. Then, we consider the item of nonuniqueness of
the \asQ definition, and the related question of coupling constant
and IC transformations. We discuss this issue in terms of coupling
transformations similar to the renormalization--scheme ones in the
weak coupling limit, but, generally, more involved and mass
dependent. Here, the important feature is {\it the vertex
dependence\,} that is typical of QFT models with gauge
symmetry.\par

 Further on, in Section 4, we return to the item of correspondence
between different IR behavior of \asQ. In particular, we give few
examples of transformations that are singular in the IR region
and, in some cases, results in the \asQ behavior in a similar way
to the ones discussed in Section 2. \par

 Our conclusion is that the question of the IR behavior of the effective
QCD coupling \asQ (and of propagators) is not a well--defined one.
The results of different groups, formulated in the momentum
representation, should not be compared directly with each other.
Calculation of hadronic characteristics remains as a reasonable
criterion for comparison of different lattice simulation schemes.

\section{Nonperturbative results in the IR region\label{s2}}
  Here, we give a short overview of nonperturbative results obtained
by different groups in calculating the effective coupling \asQ
behavior in the IR region by numerical simulation on lattice and
by approximate solution of the truncated Dyson--Schwinger
equations (DSEs). \par

 Each of these groups uses its own definition for \asQ, compatible
with the asymptotic freedom regime in the UV region, and obtains
its own result in the IR domain.

\subsection{T\"ubingen: gluons and ghosts}
 We start with a description of recent results obtained by the
T\"ubingen group. This group activity (for a review see paper
\cite{as01}) proceeds in parallel along two lines: solving the
truncated Dyson--Schwinger equations and the lattice simulation of
Euclidean 4-dimensional functional integral for quantum
gluodynamics. Thus, this group limits itself to the gluon--ghost
sector of QCD that turns out to be convenient for the DSEs
truncation in the Landau gauge. Here, the gluon--ghost vertex
renormalization function $\,\Gamma\,$ due to the Ward--Slavnov
identities can be represented via analogous functions of the gluon
$\,Z\,$ and ghost $\,G\,$ propagators. At the same time, in this
gauge, $\,\Gamma\,$ drops out from the invariant coupling defined as
\beglab{a-tub}
\albar_{Tu}(Q^2)=\as Z(Q^2)\,G^2(Q^2)\,.\eeq
Due to this, only two propagators enter into the properly truncated
DSEs.
\medskip

\underline{\sf Lattice simulations} for propagators reveal
\cite{fa02,far02} a rather specific IR behavior. The gluon function
$\,Z(Q^2)\,$ --- the transverse amplitude of the gluon propagator ---
passes through maximum at $\,Q=\sqrt{Q^2}\simeq 1\ \GeV\,$ and then
goes to zero approximately as $\,Q\,,$ the ghost one
$\,G(Q^2)\,$ is monotonous and has a power IR singularity close to
$\,Q^{-1/2}\,,$ while the product (\ref{a-tub}) tends to a finite value.
This picture is supported by
\medskip

\underline{\sf DSEs solving} that yields the power IR behavior
for the propagators
$$
G(Q^2)\,\simeq(Q)^{-\kappa}\,;\quad Z(Q^2) \,\simeq (Q^2)^{\kappa}
\quad\mbox{with}\quad \kappa=0.595$$
and finite limiting value for the invariant coupling
$$\albar_{Tu}(0)\simeq 2.97(=8.92/N_c)\,.$$

\subsection{Paris group, 3-- gluon vertex\label{ss2.2}}
 Here, the invariant QCD coupling function is constructed on the basis
of the three--gluon vertex with the nonsymmetric MOM subtraction in
the Landau gauge for the QCD model with {\sf two quark flavors}. In this
particular MOM scheme the invariant coupling
\beglab{a-par}
\albar_P(Q^2)=\as\tilde{\Gamma}^2(Q^2) Z^3(Q^2)\quad ; \quad
\tilde{\Gamma}(Q^2)\equiv\Gamma_{3gl}(Q^2,0,Q^2)\,\,,\eeq
according to lattice simulations, obeys a very peculiar behavior in
the IR region. It passes through the maximum at $Q\simeq 1 - 1.5\ \GeV\,$
and then quickly approaches zero at $Q\simeq 0.5\,\GeV\,$ -- see
Fig.3 from \cite{b1-171}.

The data here are not sufficient to define the IR limiting regime.
Nevertheless, in our opinion, the power decreasing
$\simeq (Q^2)^{\nu};\,\nu\gtrsim 2\,$ is not excluded.

\subsection{``Transoceanic team", gluon--quark vertex\label{ss2.3}}
This team\footnote{The group includes the authors from various centers
from Australia, Great Britain and continental Europe.} considers (for
a recent review see \cite{sk02}) the QCD model
with {\sf two massive quarks.} It defines its invariant coupling
$\,\gbar_{\rm TO}\,$ on the basis of the gluon--quark vertex in the
particular MOM scheme with gluon momentum equal to zero. It reads
\beglab{a-desy}
\gbar_{TO}(Q^2)=\as\Gamma_{TO}(Q^2)Z^{1/2}(Q^2)S(Q^2)\quad ;\quad
\Gamma_{TO}(Q^2)\equiv\Gamma_{q-gl}(0;Q^2,Q^2)\,,\eeq
with $\,S\,,$ the quark propagator amplitude. The results for
$\,\gbar_{\rm TO}\,$ obtained by lattice simulation, qualitatively, are
close to the ``Paris group" ones -- see Fig. 4 in paper \cite{skw01}
or Figs. 8 and 9 in \cite{sk02}. However, the IR asymptotics of IC,
evidently, has a different power
$$\,\gbar(Q)\simeq(Q^2)^{\mu/2}\,;\quad \albar
\simeq(Q^2)^{\mu}; \quad \mu\lesssim 1\,.$$

\subsection{``ALPHA" --- The Schr\"odinger functional\label{ss2.4}}
 The ``ALPHA" group considers the QCD model with {\sf two massless flavors.}
It uses the {\it Schr\"odinger functional} (SF) defined in the Euclidean
space--time manifold in a specific way: all three space dimensions are
subject to periodic boundary conditions, while the ``time" one is singled
out --- the gauge field values on the ``upper" and ``bottom" edges differ by
some phase factor $\exp{(-\eta)}$. Then, the renormalized coupling is defined
via the derivative $\Gamma'=\partial\Gamma/\partial\eta\,$ of the leading
part of the effective action
$$ \Gamma=\alpha^{-1}\Gamma_0+\Gamma_1+\as\Gamma_2+\dots $$
as (cf. eq.(2.43) in ref.\cite{lu92} and (8.3) in ref.\cite{lu99}) a
function {\it in the coordinate representation}
\beglab{al92lu}
\albarsf(L)=\Gamma'_0/\Gamma'\eeq
with $L\,$ being the spatial size of the above--mentioned manifold. \par

 To follow the $\albarsf(L)\,$ evolution, a special trick was used
 \cite{lu99}. The ``step scaling function"
$$\sigma(\albarsf(L))=\albarsf(2L)\,$$
 has been introduced.
For explicit implementation of $\sigma\,,$ the beta--function is
necessary. On the other hand, numerically, this function can be defined
from lattice simulations by comparing the results for lattice $L\,$
with $2L\,$ --- see Fig.16 in ref.\cite{lu99}. Meanwhile, fresh results
\cite{alf01b} reveal the steep rise of the SF running coupling in the
region $\albarsf\simeq 1\,$ and $L\to\infty\,.$ Here, the analytic fit
for the numerically calculated behavior of \albarsf has an exponential
form
\beglab{sf-exp}
\albarsf(L)\simeq e^{mL}\quad\mbox{with}\quad m\simeq 2.3/L_{\rm max}
\eeq
and $L_{\rm max}\,,$ a reference point in the region of sufficiently
weak coupling, indirectly defined via the condition
$\albarsf(L_{\rm max})=0.275\,. $
 For a physical discussion of the IR (i.e., at $\,Q^2\to 0\,$) QCD
behavior, in the papers of ALPHA group the
``usual quantum--mechanical correspondence" $\,L=1/Q\,$ is used.

\section{Mass--dependent coupling function\label{s3}}
 Renormalization schemes with scale parameter  $\,\mu\,$ coinciding
with subtraction momentum --- the so--called MOM schemes --- are
singled out from a formal point of view. For these schemes the condition
of the invariant coupling normalization is of a simple form
\beglab{canon}                                                         
\left.\gbar\right|_{Q^2=\mu^2}=g_{\mu}\,.\eeq
In a more general case\footnote{In this section we shall use a general
notation $\,g\,$ for a coupling constant along with a more specific one
$\,\alpha=g^2/4\pi\,$ commonly adopted in gauge theories.},
that is relevant to minimal subtraction schemes, in particular to
the \msbar one, this condition contains the ``normalization function"
\beglab{necanon}
\left.\gbar\right|_{Q^2=\mu^2}=N(g)\,,\eeq                          
that can be dependent on mass(es).\par

 The normalization (\ref{canon}) corresponds to the simplest functional
equation (FE) for the invariant coupling, that in the massless case is
of the form
\beglab{feqm0}                                                      
\gbar(x;g_{\mu})=\gbar\left(\frac{x}{t}; \gbar(t;g_{\mu})\right)\quad,
\quad x=\frac{Q^2}{\mu^2}\,\,.\eeq

 Here, the condition (\ref{canon}) follows from the RG functional
equation (\ref{feqm0}). We shall refer to (\ref{canon}) as to the
``canonical"  normalization condition. Meanwhile, for the case
(\ref{necanon}), the FE is of a more involved form --- see below
eq.(\ref{ct12}) depending on the mass argument $\,m\,$ as well.

 \subsection{Massive case}
 Now, we consider a more general case that takes into account an
important generalization: the \gbar dependence of mass(es) of particle(s).
Here, an invariant coupling acquires one more argument that can be
chosen (see, \eg ref.\cite{rg-kniga}) as a ratio $\,m^2/\mu^2=y\,,$ and
the normalization condition for an invariant coupling
\beq
\gbar_N(1,y;g)= N(y,g)\, \label{Gnorm}\eeq                       
generally differs from the simplest, ``canonical", form (\ref{canon}).
It contains a function $N\,$ depending on two arguments. \par

 Note here that we have introduced a new special notation $\,\gbar_N\,$
for the coupling function with the general normalization condition
(\ref{Gnorm}) to distinguish it from a coupling function normalized to
$\,g\,.$  In what follows, we use the {\sf term ``invariant coupling"
(IC) for the function $\,\gbar_N\,$ with the nontrivial normalization
function $N\neq g\,$ and the term ``effective coupling" (EC) for  the
function \gbar with (\ref{canon}).}\par
\smallskip

  The effective massive coupling satisfies\cite{bsh55rg1} a rather
simple functional
\beglab{feq}                                                  
\gbar(x,y; g)=\gbar\left(\frac{x}{t},\frac{y}{t},\gbar(t,y; g)\right)
\,; \quad x=\frac{Q^2}{\mu^2}\,,\,\,y=\frac{m^2}{\mu^2}\,,\eeq
and differential equations
\beglab{ct9}                                                   
x\frac{\partial\gbar(x,y; g)}{\partial x}=\beta\left(\frac{x}{y},
\,\gbar\right)\,,\quad\mathcal{X}_{\beta}\cdot\gbar(x,y;g)=0\,\eeq
with
\beglab{ct10}
\mathcal{X}_{\beta}\equiv\left\{x\frac{\partial}{\partial x}+
y\frac{\partial}{\partial y}-\beta(y,g)\frac{\partial}{\partial g}
\right\}\,;\quad\beta(y,g)\equiv \left.\frac{\partial\gbar(\xi,y;g)}
{\partial \xi}\:\right|_{\xi=1} \,.\eeq
 The last two equations (\ref{ct9}) and (\ref{ct10}) are generalizations
of massless equations, and turn to the latter in the limit $\,y\to 0\,.$
  We define a special notation $\,\mathcal{X}_{\beta}\,$ for the
particular infinitesimal (Lie) group operator (that in the given context
was first introduced by Ovsyannikov\cite{oves} in 1956). Its superscript
refers to the coordinate of the last term. In what follows we shall
refer to $\mathcal{X}\,$ as to {\it the Lie--Ovsyannikov (LO) operator.} \par

  The effective coupling $\,\gbar\,$ by definition satisfies the canonical
normalization condition
\beglab{canon-m}\gbar(1, y; g)=g\,,\eeq                         
while other RG one-argument functions, like propagator amplitudes, at
$\,Q^2=\mu^2\,$ are normalized to unity.\par
\smallskip

 At the same time, the functional equation for the IC is of a more
 involved form
\beq\label{ct12}                                         
\gbar_N(x,y;g)=\gbar_N\left\{\frac{x}{t},\frac{y}{t};
\:n\left[\frac{y}{t},\gbar_N(t,y;g)\right]\right\}\,\eeq
that reflects the group composition property and corresponds to the
normalization condition (\ref{Gnorm}). Here, $n\,$ is the function
reverse to $N\,$ with respect to the second argument.\par

 By differentiating (\ref{ct12}) in two various ways one obtains
 the differential equations
\beq\label{ct13}
{\partial\gbar_N(x,y;g)\over\partial\ln x}=B\;\left({y\over x}\:,
\gbar_N(x,y;g)\right)\,,\quad\mathcal{X}_{\beta}\cdot\gbar_N(x,y;g)=0\eeq
 with {\small
\beq\label{b-rel}
B(y,g)=\left.{\partial\over\partial\xi}\gbar_N\left\{\xi,y;n(y,g)
\right\}\right|_{\xi =1}\, ;\quad\beta(y,g)=\left.\left\{B(y,\gamma)
{\partial\over\partial\gamma}- y\frac{\partial}{\partial y}\right\}
n(y,\gamma)\right\vert_{\gamma= N(y,g)}.\eeq  }                 
 Note here that for IC we have two {\it different\/} beta--functions
$B\,$ and $\beta\,$ entering into eqs.(\ref{ct13}) and related by the
second of equations (\ref{b-rel}).\par
\medskip

Meanwhile, it turns out to be possible to treat
invariant coupling $\,\gbar_N\,$ as an effective
coupling function corresponding to some other coupling
constant $\gamma \equiv N(y,g)\,.$
 Indeed, by introducing a new coupling function
\beq\label{gam-bar}
\bar{\gamma}(x,y;\gamma)=\gbar_N\{x,y; n(y,\gamma)\}\,,\eeq      
we find that it satisfies the group functional equation of a
simple type (\ref{feq}), the differential eqs. analogous to (\ref{ct9})
with the LO operator $\,\mathcal{X}_B\,$ with the same generator
$\,B(y,\gamma)\,$ and can be considered as EC corresponding to the
coupling constant $\,\gamma\,.$ \par

  On the other hand, if we define a new auxiliary function
$\,\tilde{g}\,$ by the relation
\beq\label{ct19}
\tilde{g}(x,y;g)=n\left\{\frac{y}{x}\,,\,\gbar_N(x,y;g)\right\}=n\left\{
\frac{y}{x}\,,\,\bar{\gamma}\left(x,y;N(y,g)\right)\right\}\,, \eeq                   
it will also satisfy the simple functional equation (\ref{feq})
and can be treated on the equal footing with EC $\,\bar{\gamma}\,$.
 Both group differential eqs. for $\,\tilde{g}\,$ contain the same
generator $\beta\,.$ We shall refer to it as to the ``group effective
coupling" to stress that, generally, it is not related to IC by the
Dyson transformation.

\subsection{Mass dependent transformations\label{ss3.2}}
 Consider now, in the general mass dependent case, the transformation
from the coupling constant $\gamma\,$ and IC $\bar{\gamma}\,$ to the
new coupling constant $g\,$ and EC \gbar. The transition from popular
massless \msbar to the above--mentioned massive \msbar can be
represented as a sequence of such transformations. Let, without loss of
generality, coupling constants $g\,$ and $\gamma\,$ be connected by
$\gamma=N(y, g)\,,\quad g=n(y,\gamma)\:.$
Then, according to eq.(\ref{ct19}), the effective couplings will
be related by
\beq\label{ct18b}
\bar{\gamma}(x,y;\gamma)=N\{y/x,\gbar(x,y; g)\}\eeq
and the corresponding group generators by (\ref{b-rel}).

 In other words, it is possible to consider the transition
\beq g \to \gamma = N(y, g)\label{s.6}\eeq
as a coupling constant transformation related to the change of the
renormalization prescription. That is, starting with the set
$\{g, \gbar(x,y;g), \beta (y,g)\}$
\noindent we can move to  another one
$\{\gamma, \bar{\gamma}(x, y;\gamma), B(y,\gamma)\}$
by the transformations (\ref{s.6}) and also by (\ref{ct18b})
represented in the form
\beq\label{s.7}
\gbar\to\bar{\gamma}(x,y;\gamma)=N\left\{\frac{y}{x}\,,\gbar
\left[x,y; n\left(\frac{y}{x}\,,\gamma \right)\right]\right\}\eeq
that is a composition of (\ref{gam-bar}) and (\ref{ct19}) with the
generators $\beta\,$ and $B\,$ related by Eq.(\ref{b-rel}).\par

  Correspondingly, in the group FEq., for an RG covariant function
$s(x,y;g)\,$ (like, \eg, propagator amplitude) it is necessary not
only to take into account relation (\ref{ct19}) between IC $\gbar_N\,$
and EC $\tilde{g}\,$
\beq\label{17}
s(x,y;g)=\frac{s(t,y;g)}{s(1,y,\tilde{g}(t,y;g))}\,\cdot s\left(
\frac{x}{t}\:,\,\frac{y}{t}\:;\,\tilde{g}(t,y;g)\right)\eeq                       
\noindent but also normalization of $s\,$ itself
$$s(1,y;g) = S(y,g) \neq 1\,. $$

The related differential equations are of the form
\beq\label{s.9}
\frac{\partial\ln s(x,y; g)}{\partial\ln x}=\gamma\left[{y\over x}\:,
\tilde{g}(x,y,g)\right]\,;\quad\mathcal{X}_{\beta}\cdot\ln s(x,y;g)=
\tilde{\gamma}\left(\frac{y}{x}\:,\,g\right) \eeq
with {\small
\beq
\gamma(y, g)=\left.\frac{\partial}{\partial t}\ln
s(t,y;g)\right|_{t=1}\,\quad\mbox{and}\quad
\tilde{\gamma}(y,g)=\gamma(y,g)+\left\{y\frac{\partial}
{\partial y}-\beta(y,g)\frac{\partial}{\partial g}
\right\}\ln S(y,g)\,. \label{s.12}\eeq }
In a given gauge QFT model with one coupling, both EC and IC,
besides the renormalization scheme, are specified by the choice of the
vertex and the way of its subtraction.

\subsection{The ``vertex" dependence\label{ss3.3}}
 In the gauge theories for the mass dependent case in the
definition of effective coupling there appears a new
specific degree of freedom related to the existence of
several Lagrangian structures with the same coupling
constant. Here, in defining IC, different vertices can be
used. This means that in such a case we have various possibilities for
defining IC {\it in the same RS}. The transition between diverse ICs is
described by transformations just considered.   \par

To illustrate, take, e.g., a few dressed QCD vertices, the 3--gluons
$\,\Gamma_{3gl}\,,$ the gluon--ghost $\,\Gamma_{gl-gh}\,$ and the
gluon--quark $\,\Gamma_{gl-q}\,$ ones.

 Generally, each of them could be used for the QCD
invariant coupling defining, \eg
\beglab{verts}
\albar_{3gl}=\as\Gamma_{3gl}^2\:Z^3\,,\quad\albar_{gl-gh}=\as
\Gamma_{gl-gh}^2\:Z\,G^2\,,\quad\albar_{gl-q}=\as\Gamma_{gl-q}^2\:Z\,S^2\,\eeq
with $\,Z,\, G\,$ and $\,S\,$ --- gluon, ghost
and quark propagator scalar amplitudes. \par

  In the UV massless limit, all definitions are, in a sense, equivalent:
they coincide in the \msbar scheme and could be ``slightly different" in
MOM schemes due to various definitions of the ``reduced" vertex functions
presented as functions of one space--like argument $\,Q^2\,.$\par

However, in the massive case they could be drastically different in the IR
region due to diverse dependence on the light quark masses. As it is well
known, the quark (fermionic) propagator as well as the gluon--quark vertex
are singular on the quark mass shell. For the light quarks this leads to the
singular IR behavior.

The gluon--ghost vertex and related coupling $\,\albar_{gl-gh}\,$ seems to
be less sensitive to the mass effects. Indeed, at the one--loop level,
quark mass effects can contribute to it only via the gluon propagator
factor $Z\,.$ Meanwhile, the polarization loop is free of afore--mentioned
singularity.
Two others ICs, $\albar_{3gl}\,$ and $\albar_{gl-q}\,$ are more sensitive.
In each of them, besides three propagator amplitudes, the mass effects
come also via the vertices $\Gamma_{3gl}\,$ or $\Gamma_{gl-q}\,.$ \par

  Quite probably, just to this there corresponds a more ``quiet" variant of
the IR behavior for $\,\albar(Q^2)\,$ obtained by the T\"ubingen group.

\section{The model coupling transformations\label{s4}}
 Here, we present some model transformations of the coupling constant and
related IC transformations, including the mass--dependent ones. It is shown
that an appropriate \as transformation could drastically change the properties
of the corresponding running coupling as a function of $\,Q^2\,$ in the IR
region.

\subsection{Examples from APT of massless transformations\label{ss4.1}}
 Let us start with examples induced by the {\it Analytic Perturbation
Theory} (APT) that has been devised recently \cite{prl97,pl98} to clean
out the perturbative QCD (in a ``bloodless" way --- by imposing the
K\"allen--Lehmann analyticity) of unphysical singularities like
the Landau ghost pole. \par

In the APT, the transition from the usual invariant \msbar coupling
constant \as to the Minkowskian $\alpha_M\,$ and Euclidean $\alpha_E\,$
ones can be treated \cite{epjc} as a coupling transformations similar
to a change of the renormalization scheme (RS). At the one--loop case
\beglab{trns1-29}
\as\to\alpha_M(\as)=\frac{1}{\pi\beta_0}\arccos\frac{1}{\sqrt{1+\pi^2
\beta_0^2\as^2}}\,=\frac{1}{\pi\beta_0}\arctan(\pi\beta_0\as)\,,\eeq
\beglab{altrns2-30}
\as\to\alpha_E(\as)=\as+\frac{1}{\beta_0}\left(1-e^{1/\beta_0\as}
\right)^{-1}\,, \eeq

 Here, the first transition ``looks quite usual"\footnote{The same
function of the initial coupling constant $F(g)\sim \arctan(g)\,$ appeared
\cite{walter} in the exact solution of the two--dimensional Thirring model.}
as $\alpha_M\,$ can be expanded in powers of \as (like in the usual RS
transformation), while the second one in the weak coupling case is close to
the identity transformation as far as the nonperturbative term
$~\simeq e^{-1/\beta_0\as}$ leaves no ``footsteps" in the power expansion.\par

If one starts with the common one--loop $\albarsQ\sim 1/\ln(Q^2/\Lambda^2)\,$
with its first order pole at $Q^2=\Lambda^2\,,$ then, as a result of
substitution of (\ref{trns1-29}) and (\ref{altrns2-30}) into the RG
differential
equation and its integration, one arrives at the ghost--free expressions
\beglab{ic5}
\tildal(Q^2)=\left.\frac{1}{\pi\beta_0}\arccos\frac{L}{\sqrt{L^2+\pi^2}}
\right|_{L>0}=\frac{\arctan(\pi/L)}{\pi\beta_0}\,;\quad L=\ln\frac{Q^2}
{\Lambda^2}\,,\eeq
\beglab{ic6}
\alphan(Q^2)=\frac{1}{\beta_0}\left[\frac{1}{L}- \frac{\Lambda^2}{Q^2-
\Lambda^2}\right]\,\,\eeq
that are well known in the APT. \par

The transformation functions $\alpha_M(\as)\:$ and $\alpha_E(\as)\,$
obey important properties. They tend to \as in the weak coupling limit
\par $$\quad \alpha_i\to\as\quad \mbox{at}\quad\quad\as \ll 1\,
\phantom{~~~~(C)}\eqno{[AF]}$$\par and \par \hspace{53mm}
\underline{are finite at $\as=\infty\,.$} \hspace{56mm}[GhF] \par
Besides, they are finite
$$ \alpha_i \to 1/\beta_0 \quad\mbox{in the limit} \quad\as\to -0\,.
\phantom{~~~~(C)} \eqno{[IRf]}$$
 \smallskip

 The first property [AF] provides a correspondence with the weak coupling
limit with its asymptotic freedom property. The second one [GhF] reflects
the absence of ghosts and the [IRf] one relates to the finite IR limit.

  This last feature is absent in the transformation
\beglab{trns3n}
\as\to\alpha_{\rm N}=\as\left(1-e^{-1/\beta_0\as}\right)\,\eeq
that is equivalent to the one obtained \cite{nester} in the
modified APT.

Just due to the [AF], [GhF] and [IRf] properties, both APT ICs
(\ref{ic5}) and (\ref{ic6})\footnote{ For both $\tildal\,$ and
$\alphan\,$ the corresponding beta--functions have zero at
$\alpha=1/\beta_0$ and are symmetric under the reflection
$\left[\alpha -1/2\beta_0\right]\to-\left[\alpha-1/2\beta_0\right]\,.$
Meanwhile, the beta--function for $\tildal(s)\,$
turns out to be equal to the spectral function for $\alphan(Q^2)\,.$
This last property, that provides a peculiar realization of the
Schwinger hypothesis\cite{schw}, is valid \cite{epjc,tmp01} outside
the one--loop approximation.} \medskip

(x) enjoy the Asymptotic Freedom, \par
(xx) are free of ghost singularities at $Q^2=\Lambda^2;$
\par (xxx) have a finite IR limit at $Q^2=0\,.$ \par
\smallskip

 At the same time, the last transformation (\ref{trns3n}) satisfying only
[AF] and [Ghf], yields an expression obeying the singular IR behaviour
\beglab{ic7n}
\albar_{\rm N}(Q^2)=\frac{Q^2-\Lambda^2}{\beta_0 Q^2\ln(Q^2/\Lambda^2)}\,.\eeq
It has an ``extra $Q^{-2}\,$ factor" that, as some people believe,
relates to the linear growth of the interquark potential. \par

  Consider one more massless transformation
\beglab{trns4sf}
\as \to \alpha_{\rm SF}=
\as\frac{e^b-\exp\left\{b e^{-1/2\beta_0\as}
\right\}}{e^b-1}\,;\quad b=\frac{m}{\Lambda}\,,\eeq
leading to the expression
\beglab{ic8sf}
\albar_{\rm SF}(Q^2)= \frac{1}{\beta_0
\ln(\frac{Q^2}{\Lambda^2})}\cdot \frac{e^b-
e^{b\frac{\Lambda}{Q}}}{e^b-1}\,\eeq
with the exponential singularity $\alpha_{SF}\sim e^{M/Q}\,$ as $\,Q\to 0\,.$
It reminds an analytic approximation (\ref{sf-exp}) of recent ALPHA
group numerical results transposed (see, \eg Fig.3 in Ref.\cite{alf01b}) to
the IR momentum region with the help of ``quantum--mechanical correspondence"
$\,L\to 1/Q\,.$ Such an essential singularity contradicts the
K\"allen--Lehmann representation. Note also that expression (\ref{trns4sf}),
in turn, obeys an essential singularity at $\,\as\to +0\,.$ \par
\medskip

 From the given examples, there follows a simple ``rule
of correspondence" between transformations (\ref{trns1-29}) ---
(\ref{trns4sf}) and resulting expressions (\ref{ic5}) --- (\ref{ic8sf})
for invariant couplings: \\
\begin{quote}
{\it under the coupling constant transition $\as \to \alpha_i= f_i(\as)$
the invariant coupling function transformation is
\beglab{icft}
\albars(Q^2) \to \albar_i(Q^2)=f_i(\albarsQ) \eeq }
\end{quote} \smallskip

\subsection{Massive transformation\label{ss4.2}}
  Consider now the mass--dependent transformation of a coupling constant.
Generally, according to (\ref{s.6}), it involves a function of two variables
\beglab{38}
\as \to \alpha^* = N(y,\as)\,.\eeq
 Here, the IC transformation looks like
\beglab{37}
\albars\to\bar{\alpha}^*(x,y;\alpha^*)=N\left\{\frac{y}{x},\albars
\left[x,y;n\left(\frac{y}{x},\as\right)\right]\right\}\,.\eeq

 The r.h.s. of the last relation, in distinction to (\ref{icft}), contains
 the variable $\,y/x=m^2/Q^2\,$ that can influence the IR behavior.\par

  To illustrate, take a model expression
\beglab{m1trans}
N(y,\alpha)=\alpha_M(\as)\left(\frac{1-y}{1+y}\right)^{c\,\alpha_M(\as)}\,,\eeq
that yields
\beglab{m2trans}
\bar{\alpha}^*(Q^2)=\tildal(Q^2)\left(\frac{Q^2-m^2}{Q^2+m^2}
\right)^{c\,\tildal(Q^2)}\,.\eeq
 Here, the coupling constant $\,\alpha_M(\as)\,$ and EC $\tildal(Q^2)\,$
are defined by expressions of the type (\ref{trns1-29}) and (\ref{ic5}),
or by their more involved two-- or three--loop counterparts --- see, \eg
\cite{epjc,tmp01}. Here, it is essential that $\tildal(Q^2)\,$ is free
of unphysical singularities and monotonically increases to a finite IR
limit $\tildal(0)\sim 1\,.$  The structure of mass singularity in
(\ref{m2trans}) resembles the well--known IR one (see, \cite{rg-kniga})
of the QED vertex on the mass shell. \par

 For positive $\,c\simeq 1\,$ and small quark mass $\,m\ll\Lambda\,$
values the function $\albar^*(Q^2)\,,$ defined by (\ref{m2trans}),
can be very close to the numerical results for \albars, obtained by the
Paris group (see subsection \ref{ss2.2}) and by the ``Transoceanic team"
(subsection \ref{ss2.3}).\par

\section{Discussion}
  As it has been mentioned above in Section 2, the results of various
groups for the QCD invariant coupling in the momentum representation ---
obtained by lattice simulations of the Euclidean functional integral as
well as by approximate solving of truncated SDEs --- turn out to be
quite different in the IR region. At the same time, their ``physical"
results for hadronic properties of matter seem to be more correlated. \par

 Our ``model constructions" of Section 4 demonstrate that the IR
properties essentially depend on a precise way of the IC defining. We
used a class of ICs that in the UV region correlates with the
perturbative QCD \albars coupling in the \msbar scheme.  Just this
correspondence is usually considered to be essential in lattice
simulations and analysis of SDEs solutions. \par

 In practice, these ``admissible" ICs, satisfying the condition [AF],
can correspond to diverse lattice calculations. As far as these
calculations satisfactorily describe confinement and hadronic physics,
it is reasonable to consider them as  ``physical" ones. \par

 Special attention should be paid to  EC $\gbar^2(L)\,$ of the ALPHA
group with its exponential growth $\,\sim e^{ML}$ with lattice spatial
size $L\,.$ To interpret this result in the momentum transfer
representation, one needs to be very cautious with performing the
Fourier transformation as far as the usual Tauber criterion is not
valid. This issue will be considered in  more detail elsewhere.

 To conclude, we believe that there is no direct physical sense in
 attempts to establish some ``correct IR behavior" of the perturbative
QCD invariant coupling. Any ``infrared QCD physics" like hadronic
and $\tau$ decay ones, cannot be described in terms of only pQCD notions.
Either pQCD description has to be supplemented by some additional
semiphenomenological parameters like ``effective parton
masses"\cite{jeger97,sol01,milan02} (of order of pion mass) and anomalous
vacuum averages, or some other, intrinsically nonperturbative, means,
like lattice simulations or the Dyson--Schwinger equations, should be used.
\bigskip

{\bf\large Acknowledgements} \medskip

 The author is indebted to  R. Alkofer, G.~Altarelli, B.A. Arbuzov, V.S.
Vladimirov, M. L\"usher, I.L.~Solovtsov, and R. Stora for useful discussions
and comments. This investigation has been started in the CERN Theory Division.
The hospitality of its administration is greatly acknowledged. The work was
partially supported by grants of the Russian Foundation for Basic Research
(RFBR projects Nos 02-01-00601 and 00-15-96691), by INTAS grant No 96-0842
and by CERN--INTAS grant No 99-0377.

\end{document}